\documentclass{article}

\usepackage{graphicx} 
\usepackage[utf8]{inputenc}
\usepackage[T1]{fontenc}
\usepackage{amsmath}
\usepackage{authblk}    
\usepackage{url}        
\usepackage{newpxtext, newpxmath}   

\usepackage{bm}         
\usepackage{siunitx}    
\usepackage{cite}

\usepackage[margin=24mm]{geometry}  

\newcommand{\abs}[1]{\left|#1\right|}
\newcommand{\abssq}[1]{\abs{#1}^2}

\newcommand{\signal}{S}
\newcommand{\meassignal}{\signal}
\newcommand{\modsignal}{\hat{\signal}}
\newcommand{\bgsignal}{\modsignal_{\text{bg}}}
\newcommand{\bleach}{{\eta}}

\newcommand{\std}{\sigma}

\newcommand{\varimg}{\std_\text{img}^2}
\newcommand{\varshot}{\std_s^2}
\newcommand{\varread}{\std_r^2}
\newcommand{\vartrue}{\std_t^2}
\newcommand{\varnoise}{\std_n^2}

\newcommand{\cotrue}{c_t}
\newcommand{\coshot}{c_s}

\title{\vspace{-5mm}Inline calibration of spatial light modulators in nonlinear microscopy}

\author[1]{Daniël W.S. Cox}
\author[1]{Harish Sasikumar}
\author[1]{Ivo M. Vellekoop}
\affil[1]{Biomedical Photonic Imaging Group, Faculty of Science and Technology, University of Twente, P.O. Box 217, 7500 AE Enschede, The Netherlands}

\begin{document}
\maketitle

\begin{abstract}
We present a method for calibrating the response of a phase-only spatial light modulator in nonlinear microscopy. Our method uses the microscope image itself as calibration measurement and requires no additional hardware components.
Our method is adapted to the nonlinear signals encountered in multi-photon excitation fluorescence microscopes, and works well even under low light conditions and with strong photobleaching.
\end{abstract}

\section{Introduction}
\label{sec:intro}

Spatial light modulators (SLMs) have paved the way for a plethora of new exciting applications in the past decade \cite{Yang2023ReviewLiquid}, such as aberration-corrected optical microscopy \cite{Maurer2011WhatSpatial},
wavefront shaping \cite{kubbyWavefrontShapingBiomedical2019}, modal analysis \cite{Pinnell2020ModalAnalysis}, Raman spectroscopy \cite{Sinjab2019ApplicationsSpatial}, optical trapping \cite{Woerdemann2013AdvancedOptical} and optical communication \cite{Trichili2019CommunicatingUsing}.
For multi-photon excitation fluorescence (multi-PEF) microscopy, a type of nonlinear optical microscopy, SLMs form the key to deep-tissue imaging \cite{Katz2011FocusingCompression, tang2012SuperpenetrationOptical, kubbyWavefrontShapingBiomedical2019}.

As most hardware, SLMs must be calibrated to achieve optimal performance. This means characterizing the relation between the SLM pixel gray values (which control the applied relative voltage) and the modulated phase and amplitude, i.e. the phase and amplitude response.

The response of an SLM is wavelength dependent, and can be influenced by aging, or by experimental hardware conditions, such as the temperature of the SLM during use \cite{Li2019ProgressPhase, Zhao2022HighPrecisionCalibration}. These in turn may be influenced by application-dependent parameters (e.g. incident laser power on the SLM) and ambient conditions.
These facts makes it vital to perform frequent calibration of the SLM, under regular experimental conditions inside the optical setup.

Many different calibration procedures have been described in literature \cite{Li2019ProgressPhase}.
Some methods make use of a Twymann-Green/Michelson interferometer \cite{Xun2004PhaseCalibration, Zhang2007EvaluationPhaseonly, Mukhopadhyay2013PolarizationPhase, Teng2014CompensationMethod, Dai2019CalibrationPhaseonly, Lee2023SimpleFast} or Fizeau interferometer \cite{Lu2016ImprovedMethod}.
Others make use of auto-referencing interferometry
\cite{Zhang1994SimpleMethod, Bergeron1995PhaseCalibration, Serrano-Heredia1996MeasurementPhase, Zhang2012DiffractionBased, Engstrom2013CalibrationSpatial, Reichelt2013SpatiallyResolved, Fuentes2016InterferometricMethod, Zhao2018InterferometricMethod, Gao2019SelfreferencedMultiplebeam, Remulla2019SpatialLight, Huang2020TwoShotCalibration, Zhao2022HighPrecisionCalibration, Wang2024AbsolutelyInterferometric}.
These methods determine the phase response in various ways, such as measuring the field with
phase shifting interferometry \cite{Xun2004PhaseCalibration, Mukhopadhyay2013PolarizationPhase, Lu2016ImprovedMethod, Zhao2022HighPrecisionCalibration},
measuring the phase shift by determining the displacement of interference fringes with Fourier analysis \cite{Bergeron1995PhaseCalibration, Fuentes2016InterferometricMethod, Zhao2018InterferometricMethod, Dai2019CalibrationPhaseonly, Gao2019SelfreferencedMultiplebeam}
and by directly inverting the cosine response of the interference signal \cite{Zhang1994SimpleMethod, Serrano-Heredia1996MeasurementPhase, Zhang2007EvaluationPhaseonly, Zhang2012DiffractionBased, Engstrom2013CalibrationSpatial, Reichelt2013SpatiallyResolved, Teng2014CompensationMethod, Remulla2019SpatialLight, Huang2020TwoShotCalibration, Ordonez2024CalibrationLiquid}.
All of these options were designed to reconstruct the phase response using additional hardware, or even an entire dedicated setup.
Moreover, these methods were designed for a signal that linearly depends on the intensity. Nonlinear multi-PEF signals are therefore incompatible with these methods.
Moreover, multi-PEF microscopy typically produces signals with a low signal-to-noise ratio (SNR), and the fluorophores that produce the signal are subject to  photobleaching.
These factors must be adequately addressed to achieve an accurate calibration.

Here, we present a new inline method to calibrate the phase and amplitude response of a phase-only SLM inside a multi-PEF microscope. To the best of our knowledge, this is the first SLM calibration method for multi-PEF microscopy that does not require any hardware other than the microscope itself. Our method is designed to work well even under low-light conditions and with strong photobleaching.
We demonstrate our method in a laser-scanning 2PEF wavefront shaping microscope, and compare it with a conventional offline method using a Twymann-Green-interferometer and Fourier fringe analysis.

\section{Method}
\label{sec:method}

\begin{figure}
    \centering
    \includegraphics[width=0.5\linewidth]{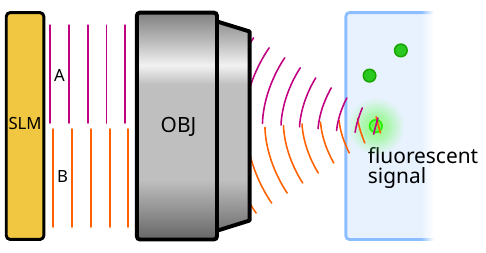}
    \caption{Principle of our measurement. SLM: Spatial Light Modulator. OBJ: Objective. The light reflects off the SLM surface, passes through the objective and converges in a focus to excite a fluorescent bead.
    The SLM is drawn at the back pupil plane of the objective. In reality, the SLM is \textit{imaged} onto the back pupil plane (see the supplemental document section 1 for a full schematic and details on the experimental setup).
    Our method splits the SLM pixels into two groups (A and B). We measure a signal for many different gray value combinations of group A and group B. The fluorescent signal depends on the interfering fields.
    }
    \label{fig:schematic-slm-groups}
\end{figure}

Our goal is to measure the phase and amplitude response of the SLM, as a function of the input parameter $g$, which, for SLMs connected to a video port, corresponds to the pixel gray value. We assume that the response $E(g)$ is uniform over the SLM, and that the SLM is conjugated to the back pupil of the microscope objective (see Fig.~\ref{fig:schematic-slm-groups}).
Our calibration method divides the SLM pixels into two groups, $A$ and $B$, with corresponding gray values $g_A$ and $g_B$ (see schematic of Fig.~\ref{fig:schematic-slm-groups}).
We detect the multi-PEF signal $\meassignal(t)$ from a scanned region of interest with a photomultiplier tube. (See supplemental document section 1 for a full schematic of our setup). We do this for combinations of all possible gray values for $g_A$ and a selection of evenly distributed values for $g_B$.
We then fit a signal model to the measured data in order to recover the response function $E(g)$.
This method relies on the interference between the fields modulated by the two halves of the SLM and how this interference manifests itself in the detected nonlinear signal.

In order to accurately model the signal, we characterize noise (see section \ref{subsec:noise}), photobleaching (see section \ref{subsec:photobleaching-model}) and nonlinear signal generation (see section \ref{subsec:signal-model}) by fitting parameterized models for each of these mechanisms.
In the last two steps, we perform weighted least-squares \cite{Strutz2011DataFitting} minimization of the following loss function:
\begin{equation}\label{eq:loss}
    \text{loss} = 
    \sum_{t}
    \frac{1}{\varnoise(t)}\left(\modsignal(t) - \meassignal(t)\right)^2
\end{equation}
where $\modsignal(t)$ is the signal calculated with our model, $\meassignal(t)$ is the measured signal, and $\varnoise(t)$ is the noise variance. Our method does not require a regularization term.

\subsection{Preprocessing}
The raw measurement data consists of images ($32\times 32$ pixels) of a small region of interest around a selected fluorescent particle.
We preprocess the data by first subtracting the mean of a dark frame (an image taken with the laser blocked) from the raw data.
We then normalize the data using the standard deviation across all measurements. This procedure minimizes the need to adjust learning parameters or initial estimates based on fluorescence intensity or signal amplifier settings.
The average pixel value of each normalized image forms our signal $\meassignal(t)$. The variance of each normalized image is used for noise analysis in section~\ref{subsec:noise}.

\subsection{Noise model}
\label{subsec:noise}

\begin{figure}
    \centering
    \includegraphics[width=0.6\linewidth]{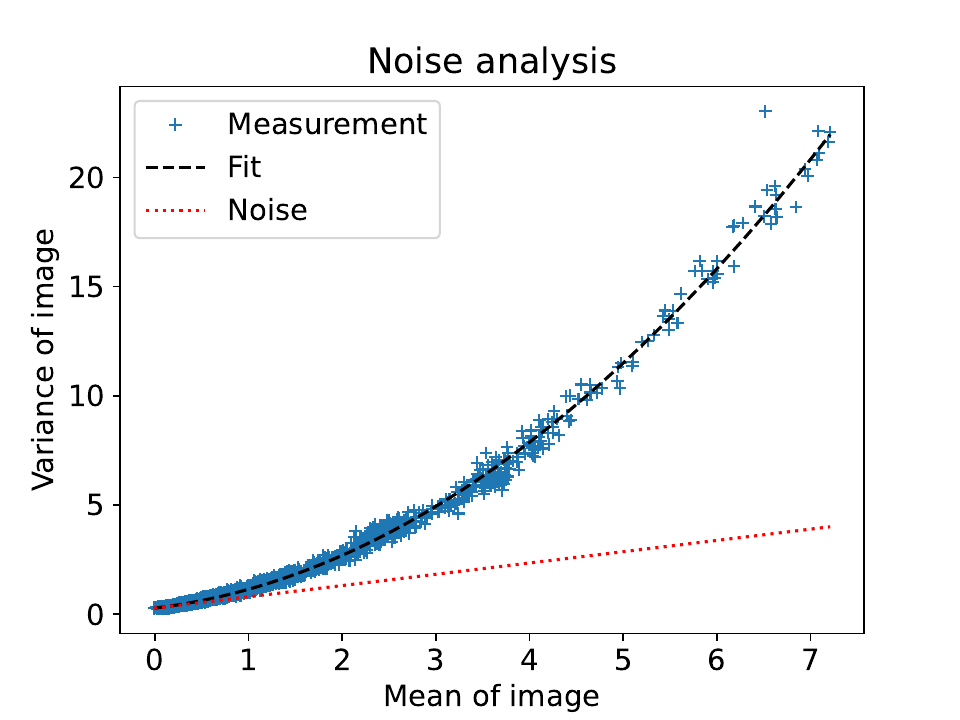}
    \caption{Image variance $\varimg(t)$ vs. image mean $\signal(t)$ for all images in the measurement sequence. Blue plusses: measurements. Black dashed curve: fitted variance model (Eq. \eqref{eq:noise-model}). Red dotted curve: estimated contribution of noise to the total variance.}
    \label{fig:noise-fit}
\end{figure}
The weighted least squares method (Eq.~\ref{eq:loss}) requires an accurate estimate of the noise variance of each measurement \cite{Strutz2011DataFitting}. To estimate the noise level for each measurement, we use a noise model that includes both read noise $\varread$ and shot noise $\varshot(t)$. 

To estimate the relative contributions for the read noise and the shot noise, we consider the pixel-to-pixel variance $\varimg(t)$ of each of the small images recorded in the experiment. We model this variance as a sum of three uncorrelated components: read noise $\varread$, shot noise $\varshot(t)$, and the variance of the `true signal' $\vartrue(t)$ (i.e. the variance of intensity distribution within the small image, independent of noise):
\begin{equation}\label{eq:varimg}
    \varimg(t) = \varread + \varshot(t) + \vartrue(t)
\end{equation}
By definition, the variance of the read noise is the same for all measurements, and the variance of the shot noise scales linearly with the measured signal \cite{Janesick2007PhotonTransfera}. The variance of the true signal scales with $\signal^2(t)$. Eq.~\ref{eq:varimg} may now be rewritten, resulting in the following quadratic function of $\signal(t)$:
\begin{equation}\label{eq:noise-model}
        \varimg(t) = \varread + \coshot S(t) + \cotrue S^2(t)
\end{equation}

Fig.~\ref{fig:noise-fit} shows the image variance  $\varimg(t)$ versus the signal $S(t)$ (i.e. the image mean) for each of the measurements. The dashed line is a least-squares fit of Eq.~\eqref{eq:noise-model}, with $\varread=0.27$, $\coshot=0.52$ and $\cotrue=0.35$. The quadratic function clearly fits the relationship between image variance and image mean very well.

With the fitted coefficients, we can estimate the noise variance $\varnoise(t)$ for each image as a function of $\signal(t)$:
\begin{equation}\label{eq:varnoise-compute}
    \varnoise(t) = \varread + \coshot \signal(t)
\end{equation}
which is represented by the red dotted line in Fig.~\ref{fig:noise-fit}. We can now use $1/\varnoise(t)$ as weighting factor for each of the measurements (see Eq.~\ref{eq:loss}).

\subsection{Photobleaching model}
\label{subsec:photobleaching-model}

\begin{figure}
    \centering
    \includegraphics[width=0.6\linewidth]{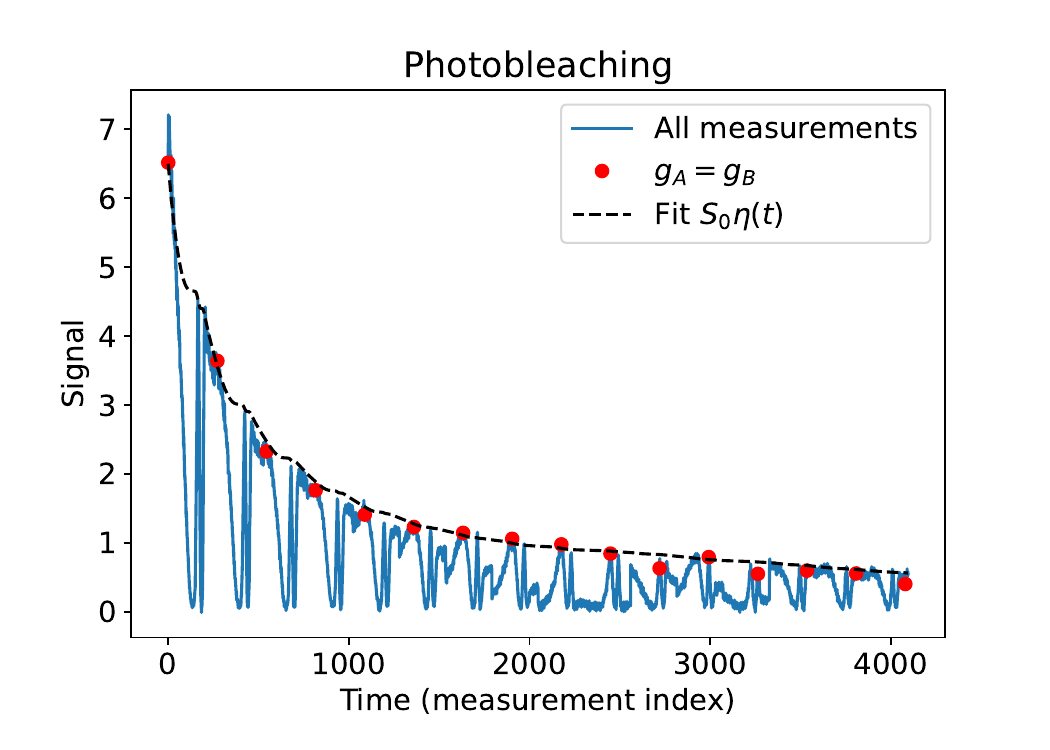}
    \caption{Solid blue line: all measured signals plotted over time (measurement index). The signal varies rapidly over time, due to varying constructive and destructive interference (see section~\ref{subsec:signal-model}). Additionally, the signal intensity decays over time due to photobleaching. The measurements took approximately 17~minutes in total (including automatically selecting a region of interest). Red dots: the measurements taken with a flat wavefront and maximum constructive interference.
    The black dashed line shows the fit through the red dots: $\signal_0 \bleach(t)$.
    }
    \label{fig:photobleaching-fit}
\end{figure}

Throughout our measurement, we vary the gray values on the SLM to produce constructive and destructive interference (see section~\ref{subsec:signal-model}). This can be observed in Fig.~\ref{fig:photobleaching-fit} as rapid variations in the signal as a function of time.
However, fluorescent dyes typically photobleach when they are excited, which reduces the signal intensity over time. 
The effect of photobleaching is clearly visible, as the signal strength decreases over time by roughly a factor of 8.

Hence, in order to accurately model our signal, we must include the effect of photobleaching.
There are several factors that complicate modeling photobleaching.
Firstly, at this point the excitation intensity is an unknown function of the SLM's gray values, meaning that the time-dependent excitation that causes the bleaching is unknown.
Secondly, different fluorophores will be excited at different rates, depending on their location in the focal volume and their orientation with respect to the polarization of the excitation light \cite{McClain1974TwophotonMolecular}.
As a result, different fluorophores will also bleach at different rates, causing the signal decay to be highly non-exponential.

We introduce the signal efficiency factor $\bleach(t)$, which describes the effects of photobleaching on the signal as a function of time $t$.
We model $\bleach(t)$ with a simple empirical model relating $\bleach(t)$ to the total signal detected so far:
\begin{equation}\label{eq:photobleach-efficiency}
    \bleach(t) =
    \exp\left(-P \!\int_0^t \meassignal(t') \,dt' \right)
\end{equation}
where $\meassignal(t)$ is the measured signal.
$P$ is the photobleaching rate of the signal.
In the supplemental document section 2, we give a derivation of Eq.~\ref{eq:photobleach-efficiency}, and show how it may be modified to account for accelerated photobleaching if needed.

In Fig.~\ref{fig:photobleaching-fit}, the red dots indicate measurements where the SLM displays a flat wavefront to produce maximum constructive interference (i.e. when $g_A(t)=g_B(t)$, see section \ref{subsec:signal-model}).
We assume that the phase-only SLM does not significantly modulate the amplitude and thus the excitation intensity must be approximately equal for all these points.
The following equation then holds (for these points only):
\begin{equation}
    \label{eq:photobleach-signal}
    \begin{split}
    \modsignal(t) & = \signal_0 \, \bleach(t) \\
    & = \signal_0 \,
    \exp\left(-P \!\int_0^t \meassignal(t') \,dt' \right)
    \end{split}
\end{equation}

where $\signal_0$ represents the unbleached signal at maximum constructive interference.
Inserting this model into Eq.~\ref{eq:loss}, and only taking into account the measurements where $g_A(t)=g_B(t)$, we performed a weighted least squares fit to find the prefactor $\signal_0$ and photobleaching rate $P$. For this fit, we used 250 iterations of the AMSGrad algorithm \cite{reddiConvergenceAdam2018}, using $\signal(0)$ as the initial guess for $\signal_0$ and initializing $P$ so that Eq.~\ref{eq:photobleach-signal} fits exactly through the first and last measurement.

Once these parameters are determined, we can estimate $\bleach(t)$ for the entire signal with Eq.~\ref{eq:photobleach-efficiency}.
It can be seen that the fitted signal (black dashed line in Fig.~\ref{fig:photobleaching-fit}) matches the envelope of the decaying signal well. As expected, $\bleach(t)$ decays steeply at times $t$ when the signal is at a local maximum.
Finally, $\bleach(t)$ is used in Eq.~\ref{eq:signal-power} in section \ref{subsec:signal-model}.

\subsection{Signal model}
\label{subsec:signal-model}

In our method, the light from the two illuminated SLM pixel groups (A and B) interferes in the focal plane (see Fig.~\ref{fig:schematic-slm-groups}). We model the intensity in the focus as 
\begin{equation}\label{eq:intensity}
    I(t) = \abssq{a E(g_A(t)) + b E(g_B(t))}
\end{equation}
where $a$, $b$ are the complex transmission coefficients from the SLM pixel groups to the focus. The gray values $g_A(t)$ and $g_B(t)$ are controlled and varied by the hardware throughout the measurement.

Although the exact behavior of multi-PEF signals can be quite intricate \cite{Lakowicz2002TopicsFluorescence, cheng2014MeasurementsMultiphoton, sinefeldAdaptiveOpticsMultiphoton2015}, the fluorescent signal strength $\modsignal$ may be approximated with a power relation \cite{Katz2014NoninvasiveNonlinear, sinefeldAdaptiveOpticsMultiphoton2015}:
\begin{equation}\label{eq:signal-power}
\begin{split}
    \modsignal(t)
    & = \bleach(t) \, I^N\!(t) + \bgsignal \\
    & = \bleach(t) \abs{a E(g_A(t)) + b E(g_B(t))}^{2N} + \bgsignal
\end{split}
\end{equation}
where $\bgsignal$ denotes the background signal and $N$ is the nonlinear order. In theory, $N=2$ when measuring a 2PEF signal with a point detector.
In practice, a slightly lower nonlinearity order $N$ is observed for our signal model (Eq.~\ref{eq:signal-power}), as the signal depends on the size and shape of the focus and the fluorescent particles \cite{Lakowicz2002TopicsFluorescence, Katz2014NoninvasiveNonlinear, sinefeldAdaptiveOpticsMultiphoton2015}.

\begin{figure*}
    \centering
    \includegraphics[width=1.0\linewidth]{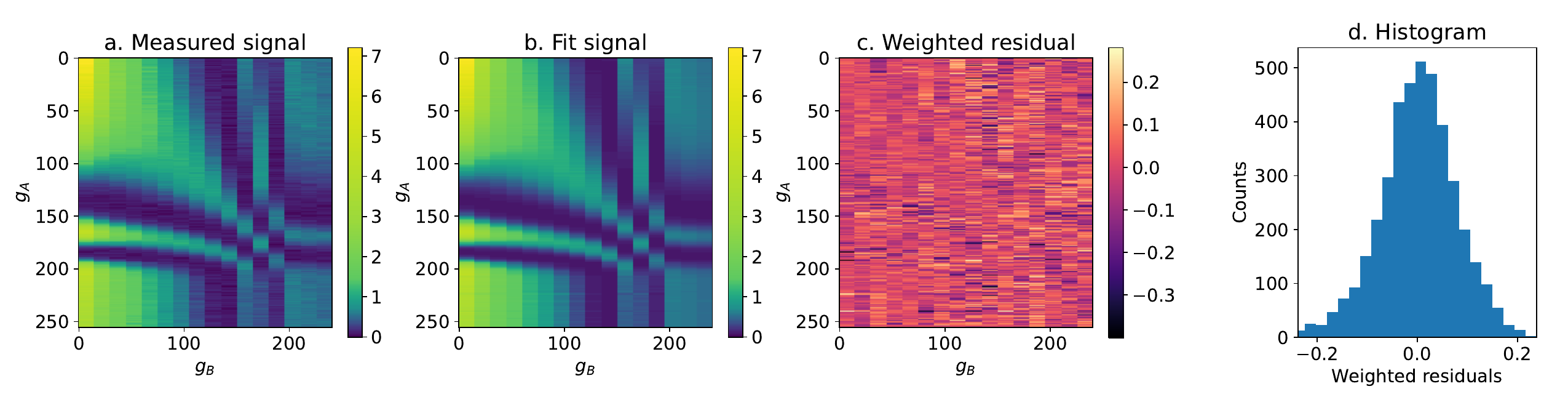}
    \caption{Signal for various combinations of gray values $g_A$ and $g_B$. a. Measured signal. b. Fit signal. c. Weighted residual for each gray-value pair. d. Histogram of the weighted residuals.}
    \label{fig:calibration-signal}
\end{figure*}

Before running the measurements, we apply the lookup table (LUT) provided by the SLM manufacturer (Meadowlark). This LUT makes the gray values correspond linearly to the voltage applied over each SLM pixel.
We measure the signal $\signal$ for combinations of all 256 gray values $g_A$ and 16 different gray values $g_B$. 
Fig.~\ref{fig:calibration-signal}a shows the measurements for each $(g_A, g_B)$ combination. The effects of constructive and destructive interference are clearly visible.
Furthermore, aside from photobleaching, the measured data appears symmetric across the $g_A=g_B$ line. Both of these facts are in accordance with our signal model (Eq.~\ref{eq:signal-power}).

We initialize $E(g)$ as a random complex function of $g$ with unit amplitude. 
We then learn $E(g)$, $a$, $b$, $\bgsignal$ and $N$ by fitting Eq.~\ref{eq:signal-power} to the measurements by minimizing the loss of Eq.~\ref{eq:loss} with AMSGrad \cite{reddiConvergenceAdam2018} in 50000 iterations. This process takes less than a minute on an office PC (Intel i7-8700 CPU). 

The results of the fit are shown in Fig.~\ref{fig:calibration-signal}b. The fit matches the measured signal (Fig.~\ref{fig:calibration-signal}a) well.
The weighted residuals $(\hat{S}(t)-S(t)) / \sigma_n(t)$ in Fig.~\ref{fig:calibration-signal}c show no systematic deviations, and the bell-shaped histogram of the weighted residuals (Fig.~\ref{fig:calibration-signal}d) shows no significant outliers. These results indicate that the weighting of residuals has been performed correctly and that the model accurately describes the experiment.

Note that we chose to fit a complex-valued field response $E(g)$ rather than a phase-only response $\phi(g)$. This approach has two advantages.
Firstly, even a phase-only SLM's field response could contain a bias due to coherent reflections off the SLM's front surface. Such a bias may also be caused by an imperfectly aligned polarization, since many SLMs only modulate the phase for one polarization component of the light. Learning $E(g)$ incorporates any bias into the response.
Secondly, although restricting the response to values with a unit modulus would leave us with fewer degrees of freedom to solve, a unit-modulus constraint by explicit parameterization can severely complicate optimization problems
\cite{Zhang2006ComplexQuadratic, Tranter2017FastUnitModulus}.

\section{Field-response results and comparison}
\label{sec:field-response}

\begin{figure}
    \centering
    \includegraphics[width=0.65\linewidth]{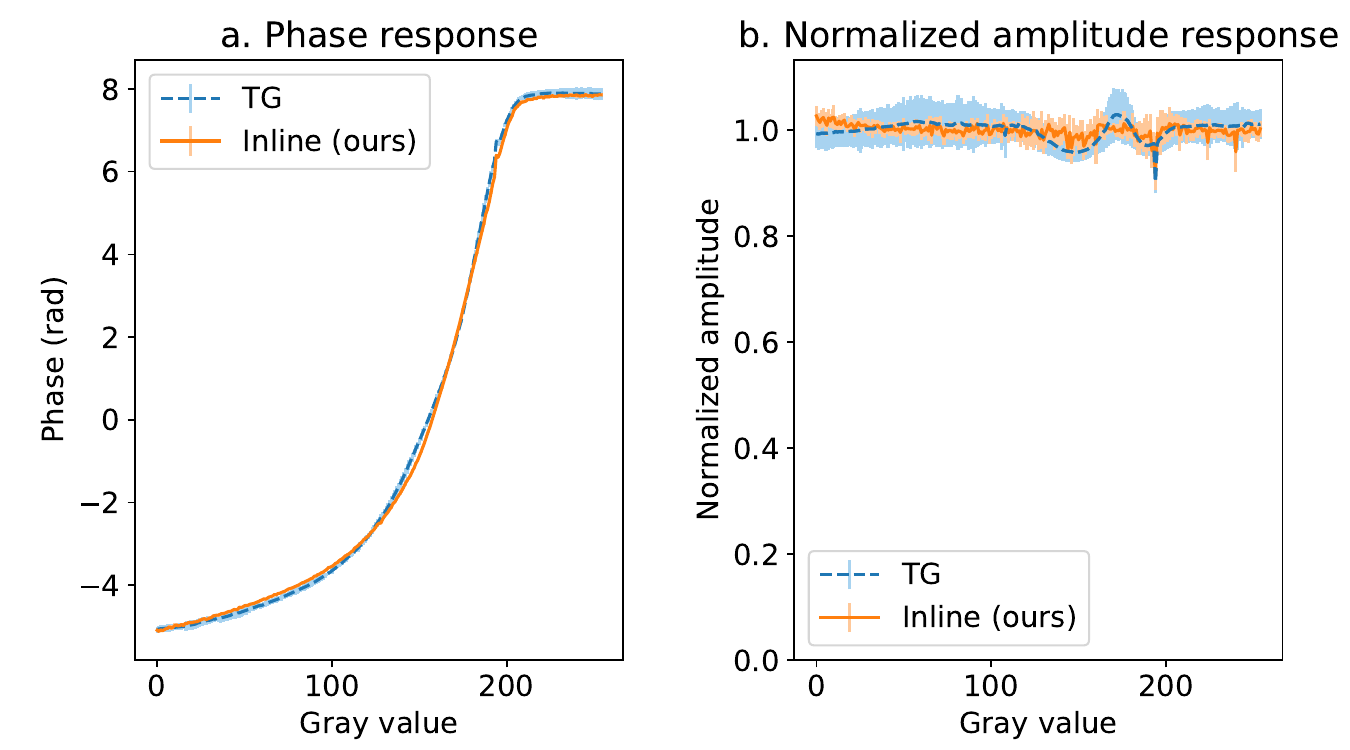}
    \caption{Comparison of (a) the phase and (b) amplitude (normalized to mean amplitude response), using our inline method (solid orange line) and a conventional method (dashed blue line).
    The conventional calibration method uses a Twymann-Green (TG) interferometer and Fourier fringe analysis.
    The conventional measurements were performed with low laser output power (\SI{0.14}{W})
    in CW mode. The inline measurements were performed with high laser output power (\SI{2.6}{W})
    in pulsed mode. The lines indicate the median and the error bars indicate the standard deviation of multiple runs (10 runs for the conventional and 9 runs for our inline method).}
    \label{fig:phase-amp-response}
\end{figure}

Fig.~\ref{fig:phase-amp-response} shows the results of our inline method. The reconstructed phase response as a function of gray value is shown in Fig.~\ref{fig:phase-amp-response}a. We repeat the method 9 times on different parts of the sample.
The line in error bars indicate the standard deviation of these results. The error bars of the phase response are very small (\SI{0.03}{rad} on average), indicating a very high precision.
Fig.~\ref{fig:phase-amp-response}b shows the normalized amplitude response with error bars of 2\% on average, which corresponds to an intensity fraction of \num{4e-4}.

For comparison, we performed a conventional calibration using Fourier analysis of the interference fringes of a Twymann-Green (TG) interferometer, (used in e.g. \cite{Dai2019CalibrationPhaseonly}). Its results are shown in Fig.~\ref{fig:phase-amp-response} with the label `TG'.
We have performed the measurements 10 times and show the median result for each gray value in Fig.~\ref{fig:phase-amp-response}. The error bars indicate standard deviation over the 10 results for each gray value. The error bars of the phase response are relatively small (\SI{0.09}{rad} on average), yet still significantly larger than for our new method.
Fig.~\ref{fig:phase-amp-response}b shows the normalized amplitude response with error bars of 3\% on average, which corresponds to an intensity fraction of \num{9e-4}.

Note that the TG measurement was performed using the Ti:sapphire laser in continuous wave (CW) to see interference fringes without exactly matching the path lengths of the arms of the interferometer. In CW mode, the stability of the laser is impaired and we observed regular shifts and jumps of the interference fringes when operated in this mode, as well as significant fluctuations in intensity. Moreover, the laser had to operate at a low output power (\SI{0.14}{W}) to enable CW mode, while our inline measurement was performed at the normal settings for operating the microscope: high laser output power (\SI{2.6}{W}) in pulsed mode.

This difference in incident laser power likely affects the temperature of the SLM. Therefore, we do not expect the phase response curves to match perfectly. Indeed, the phase response curves differs up to \SI{0.4}{rad}. Since this difference is much larger than the precision of either method and there are no systematic errors in the fit, we believe that both methods work correctly. Hence, the difference in the phase response is really present, and may be caused by the difference in laser settings (laser power and pulsed vs. continuous mode) and temperature of the SLM.

A closer examination of the amplitude response of the TG method (Fig.~\ref{fig:phase-amp-response}b) reveals a `wiggle', most noticeable between gray values $g\in[120, 200]$.
This `wiggle' occurs due to a constant bias in the field, corresponding to a fraction of the light that is not modulated.
This may be caused by a reflection off the front surface of the SLM. This effect would not play a role when the laser is operated in pulsed mode, as the pulse length (\SI{<30}{\micro m} from manufacturer specifications) is significantly shorter than a round trip through the SLM's front layer.
This bias field has an amplitude of \SI{3}{\%} of the average amplitude response, corresponding to an intensity fraction of \num{9e-4}.
We found that the wiggle disappears if we subtract the bias field from $E(g)$.

Aside from this small `wiggle', the amplitude responses of the two methods are constant (as expected for a phase-only SLM) and match within the error margin.

\section{Conclusion}
\label{sec:conclusion}
We have developed and demonstrated an inline method to calibrate the phase and amplitude response of a phase-only SLM in a multi-PEF microscope. Our method requires no additional hardware.
Our method displays precise results under the stringent constraints of low SNR and strong photobleaching, which are inherent in multi-PEF microscopy.
With this, our method makes inline SLM calibration under operational settings a possibility for multi-PEF microscopy.

\section{Data and code availability}
\label{sec:data}
This article makes use of our Python library OpenWFS \cite{vellekoop2023OpenwfsGithub, doornbos2025OpenWFSLibrary}. The code for performing the measurements and analysis is available on Github \cite{Cox2025Dedean16Inline_slm_calibration}, and our measurement data is available at \cite{Cox2025RawMeasurement}.

\bibliographystyle{unsrt}
\bibliography{main}

\end{document}


\maketitle

\section{Experimental setup}

\begin{figure}[h]
    \centering
    \includegraphics[width=1\linewidth]{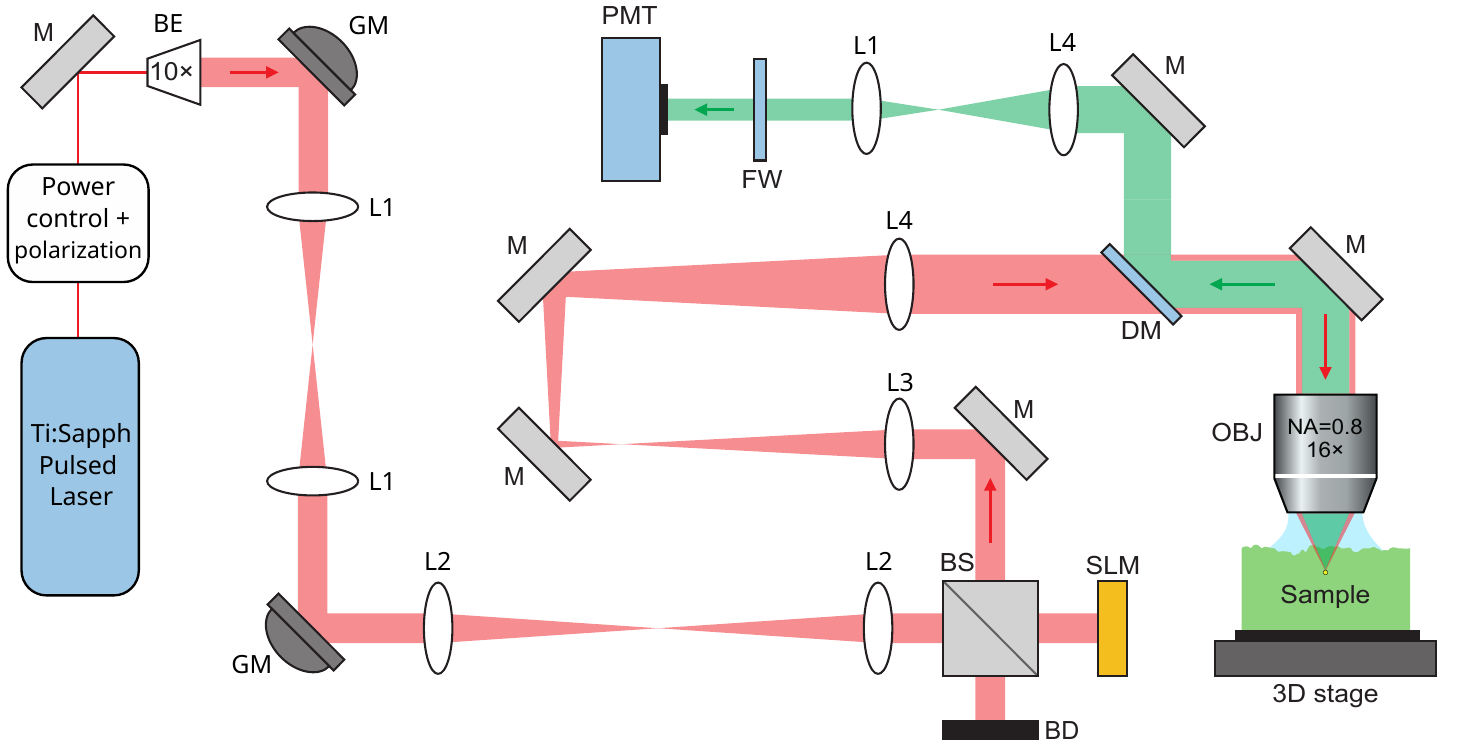}
    \caption{Schematic of the experimental setup. M: mirror, BE: beam expander, GM: galvo mirror, L: lenses with the following focal distances: L1: \SI{100}{mm}, L2: \SI{200}{mm}, L3: \SI{150}{mm} and L4: \SI{300}{mm}, BS: 50/50 beam splitter, BD: beam dump, SLM: spatial light modulator, DM: dichroic mirror, OBJ: Objective, FW: filter wheel, PMT: Photomultiplier tube. Reproduced with permission from \cite{Cox2025OrthonormalizationPhaseonly}.}
    \label{fig:2pef-microscope}
\end{figure}

Our setup is a two-photon excitation fluorescence (2PEF) laser-scanning wavefront shaping microscope. See Fig.~\ref{fig:2pef-microscope}. The SLM (Meadowlark 1920x1152 XY Phase Series) is illuminated with a linearly polarized Gaussian beam. This light is produced with a titanium-sapphire laser (Spectra-Physics, Mai Tai) with a wavelength of \SI{804}{nm}. The SLM is imaged onto the back pupil plane of a water-dipped microscope objective (Nikon CFI75 LWD 16X W). The objective forms a focus, which is employed to excite fluorescent beads (Polysciences Fluoresbrite, plain YG, \SI{500}{nm} microspheres) inside a cast PDMS sample. We detect this emitted light with a photomultiplier tube (PMT, Hamamatsu H7422-40), which produces our raw signal.

\section{Photobleaching model derivation}
In multiphoton microscopy, the generated signal is proportional to $I^N$, where $I$ is the excitation intensity.
In contrast, the exponential decay rate of the fluorophores is proportional to $I^\beta$, where $\beta$ may be slightly higher than $N$, an effect known as accelerated photobleaching \cite{Koester1999Ca2Fluorescence, Patterson2000PhotobleachingTwoPhoton, Hopt2001HighlyNonlinear, Cranfill2016QuantitativeAssessment}.

If all fluorophores are illuminated equally, we would expect the signal to be proportional to
\begin{equation}
    S_\text{naive}(t)=I^N(t) e^{-P F(t)}
\end{equation}
where we introduced the cumulative photodamage function $F(t)\equiv \int_0^t I^\beta(t') dt$, and photobleaching rate $P$. In practice, however, not all fluorophores are excited equally, and thus also do not bleach at the same rate \cite{Cranfill2016QuantitativeAssessment}.
This distribution is caused by the variation of the excitation intensity over the focal volume, as well as the distribution of orientations of the fluorophores with respect to the polarization of the excitation light, and possibly even factors like the local chemical environment of the fluorophore.

To model this distribution, we introduce $p(P)$, which describes which fraction of the observed signal originates from fluorophores with bleaching rate $P$. The total signal is now given by
\begin{equation}\label{eq:signal-bleached}
    S(t)=I^N(t) \int_0^\infty p(P) e^{-P F(t)} dP
\end{equation}
with $I(t)$ the intensity in the center of the focal volume. For convenience, we model $p(P)$ with a Gamma distribution
\begin{equation}\label{eq:decay-distribution}
    s(P) = \frac{\lambda^\alpha}{\Gamma(\alpha)} P^{\alpha-1} e^{-P \lambda}
\end{equation}
with $\Gamma$ the gamma function, and $\alpha$ and $\lambda>0$ the parameters of the distribution. We can now substitute $p(P)$ into ~\eqref{eq:signal-bleached} and evaluate the integral
\begin{equation}
    S(t)=I^N(t)\left(\frac{\lambda}{F(t)+\lambda}\right)^\alpha
\end{equation}

This expression gives the total signal $S(t)$ as a function of $F(t)$, which in turn is a function of $I(t)$. However, for our application, we want to invert this relation. For a given measurement $S(t)$, and model parameters $\alpha, \beta$ and $\lambda$, we wish to find the excitation intensity $I(t)$.

To come from $S(t)$ to $I(t)$, we first process the measured signal to compute an axillary function $S_\beta(t)$:
\begin{equation}\label{eq:S-beta-def}
    S_\beta(t)
    \equiv \int_0^t \left(S(t') \right)^\frac{\beta}{N} dt'
\end{equation}
Our model then predicts:
\begin{align}
    S_\beta &= \int_0^t \left(\frac{\lambda}{F(t)+\lambda}\right)^\frac{\alpha\beta}{N} \, I^\beta(t) \,d t' \\
    & = \int_0^{F(t)}\left(\frac{\lambda}{F'+\lambda}\right)^\frac{\alpha\beta}{N} dF'
\end{align}  
where we used a change of variables with $dF=I^\beta dt$ in the last step. We now define $\gamma\equiv\alpha\beta/N$ and evaluate the integral, and assuming $\gamma \neq 1$,
\begin{align}
    S_\beta(t) &= \frac{\lambda}{1-\gamma}\left(
    \left[\frac{\lambda}{F(t)+\lambda}\right]^{\gamma-1} -1\right)
\end{align}
which can be inverted to
\begin{equation}
    F(t) = \lambda\left[1 + \frac{1-\gamma}{\lambda}S_\beta(t)\right]^{\frac{1}{1-\gamma}}-\lambda
\end{equation}
We can thus use $S_\beta(t)$, as computed from the measurements, to find $F(t)$. Finally, we take the time derivative to recover the intensity:
\begin{equation}\label{eq:bleaching-compensated}
    I^N(t)=\left(\frac{dF}{dt}\right)^\frac{N}{\beta}=S(t) \left(1 + \frac{1-\gamma}{\lambda}S_\beta(t)\right)^{\frac{\gamma N}{(1-\gamma)\beta}}
\end{equation}
giving the excitation intensity $I(t)$ as a function of measured signal $S(t)$.

In our experiments, we find a good fit by taking $\beta=N$ and the limit $\gamma\rightarrow 1$. Consequently $\alpha=1$, which corresponds to a Gamma distribution with many fluorophores with a low bleaching rate (i.e. those illuminated far away from the focus) and few fluorophores with a high bleaching rate (i.e. those illuminated at the focus). In the limit $\gamma\rightarrow 1$, \eqref{eq:bleaching-compensated} simplifies to:
\begin{equation}
    I^N(t)= S(t) \exp\left(\frac{N S_\beta(t)}{\lambda\beta}\right)
\end{equation}
and inserting $\beta=N$ gives:
\begin{equation}\label{eq:bleaching-final}
    S(t)=I^N(t)\exp\left(-P\int_0^t S(t') dt\right)
\end{equation}
where $P \equiv 1/\lambda$. In our experiment, we have several data points where the SLM displayed a flat wavefront. Assuming that $I(t)=I_\text{flat}$ for all these measurements, we can fit $\eqref{eq:bleaching-final}$ to find the value for $P$, and the prefactor $S_0=I^N_\text{flat}$.

\bibliographystyle{unsrt}
\bibliography{supplemental}